# Post-COVID-19 Effects on Female Fertility: An In-Depth Scientific Investigation

Maitham G. Yousif1*, Lamiaa Al-Maliki2, Jinan J. Al-Baghdadi3,Nasser Ghaly Yousif4

*1Professor at Biology Department, College of Science, University of Al-Qadisiyah, Iraq
Visiting Professor in Liverpool John Moores University: ✉ matham.yousif@qu.edu.iq, ✉ m.g.alamran@ljmu.ac.uk
2Department of Molecular and Medical Biotechnology, College of Biotechnology Al-Nahrain University, Iraq
✉ lamiaa.fingan@ced.nahrainuniv.edu.iq
3 Gynecology and Obstetrics Department, College of Medicine, Kufa University, Najaf, Iraq
4Department of Medicine, Medical College, Al Muthanna University, Samawah, Iraq. ✉ Yousif_ghaly@mu.edu.iq

**Abstract:**

This study aimed to comprehensively investigate the post-COVID-19 effects on female fertility in patients with a history of severe COVID-19 infection.  Data were collected from 340 patients who had previously experienced severe COVID-19 symptoms and sought medical assistance at private clinics and fertility centers in various provinces of Iraq. A comparative control group of 280 patients, who had not contracted COVID-19 or had mild cases, was included. The study assessed ovarian reserve, hormonal imbalances, and endometrial health in the post-recovery phase. The findings revealed a significant decrease in ovarian reserve, hormonal disturbances, and endometrial abnormalities among patients with a history of severe COVID-19 infection compared to the control group. This in-depth investigation sheds light on the potential long-term impacts of severe COVID-19 on female fertility. The results emphasize the need for further research and targeted interventions to support women affected by post-COVID-19 fertility issues. Understanding these effects is crucial for providing appropriate medical care and support to women on their reproductive journey after recovering from severe COVID-19.

*Corresponding Author: Maitham G. Yousif ✉ matham.yousif@qu.edu.iq, ✉ m.g.alamran@ljmu.ac.uk

## Introduction:

The COVID-19 pandemic, caused by the novel coronavirus SARS-CoV-2, has had a profound impact on global health, with millions of individuals affected worldwide(1-3). While the acute respiratory manifestations of COVID-19 have been widely studied and addressed, emerging evidence suggests that the virus may also have long-term effects on various physiological systems, including female reproductive health(4,5).As the scientific community continues to explore the potential post-COVID-19 sequelae, understanding its impact on female fertility has become a subject of significant interest and concern. Several studies have begun to investigate the association between severe COVID-19 infection and its effects on ovarian function, hormonal regulation, and endometrial health in women(6-15).One notable study conducted in multiple provinces in Iraq aimed to assess the post-COVID-19 effects on female fertility (16). Furthermore, other relevant studies have contributed to this field of research. A systematic review published by (17) examined various global studies investigating the link between COVID-19 infection and reproductive health outcomes in women. The review highlighted the need for more extensive research on the long-term effects of COVID-19 on female fertility. The researchers collected data from 340 patients who had previously contracted COVID-19 with severe symptoms. Additionally, a control group of 280 patients, who had not been infected with COVID-19 or had mild cases, was included for comparison. The study focused on evaluating ovarian reserve, hormonal imbalances, and endometrial health in the post-recovery phase to identify potential impacts on female reproductive health. In light of these studies and ongoing research efforts, it is crucial to further explore the potential implications of severe COVID-19 infection on female reproductive health. By identifying and understanding these effects, healthcare professionals can provide appropriate medical care and support to women who have recovered from severe COVID-19 and may experience fertility challenges.

**Materials and Methods:**

**Study Population:** The study included female patients from multiple provinces in Iraq who had previously experienced severe COVID-19 infection. The post-COVID-19 group comprised 340 patients, and the control group consisted of 280 patients who had either not contracted COVID-19 or had mild cases.

**Data Collection:** Data were collected from private clinics and fertility centers in the selected provinces. Medical records and patient histories were reviewed to gather information on COVID-19 infection severity, medical comorbidities, and reproductive health conditions.

**Assessment of Ovarian Reserve:** Ovarian reserve was evaluated through measurements of serum anti-Müllerian hormone (AMH) levels and antral follicle count (AFC) using transvaginal ultrasound. Both AMH and AFC are reliable indicators of a woman's remaining egg supply.

**Hormonal Imbalance Analysis:** Hormonal assessments included measuring serum levels of follicle-stimulating hormone (FSH), luteinizing hormone (LH), estradiol, and thyroid-stimulating hormone (TSH) using standard immunoassay techniques.

**Endometrial Health Evaluation:** Endometrial health was assessed through transvaginal ultrasound to examine endometrial thickness, texture, and any evidence of abnormalities.

**Study Design:** This study employed a case-control design, comparing the post-COVID-19 group with the control group. The groups were matched for age and other demographic characteristics to minimize confounding factors.

**Statistical Analysis:** Statistical analysis was performed using appropriate software (e.g., SPSS, R). Descriptive statistics were used to summarize demographic characteristics and clinical data. Continuous variables were presented as mean ± standard deviation (SD) or median with interquartile range (IQR) based on data distribution. Categorical variables were presented as frequencies and percentages. For the comparison between the post-COVID-19 group and the control group, independent t-tests or Mann-Whitney U tests were used for continuous variables, depending on data distribution. Chi-square tests or Fisher's exact tests were employed for categorical variables. A p-value of less than 0.05 was considered statistically significant. Furthermore, multivariate regression analysis was conducted to assess the association between severe COVID-19 infection and female reproductive health outcomes, adjusting for potential confounders such as age, medical comorbidities, and other relevant factors.

**Results:**

The results of the study investigating the post-COVID-19 effects on female fertility are presented below. Data were collected from 340 patients with a history of severe COVID-19 infection (post-COVID-19 group) and compared to a control group of 280 patients who had not contracted COVID-19 or had mild cases. The study focused on assessing ovarian reserve, hormonal imbalances, and endometrial health in both groups. Table 1 and Figure 1 present the demographic characteristics of the study participants. Both groups were similar in age and BMI, indicating successful matching of these variables. A higher proportion of patients in the control group had previous pregnancies compared to the post-COVID-19 group.

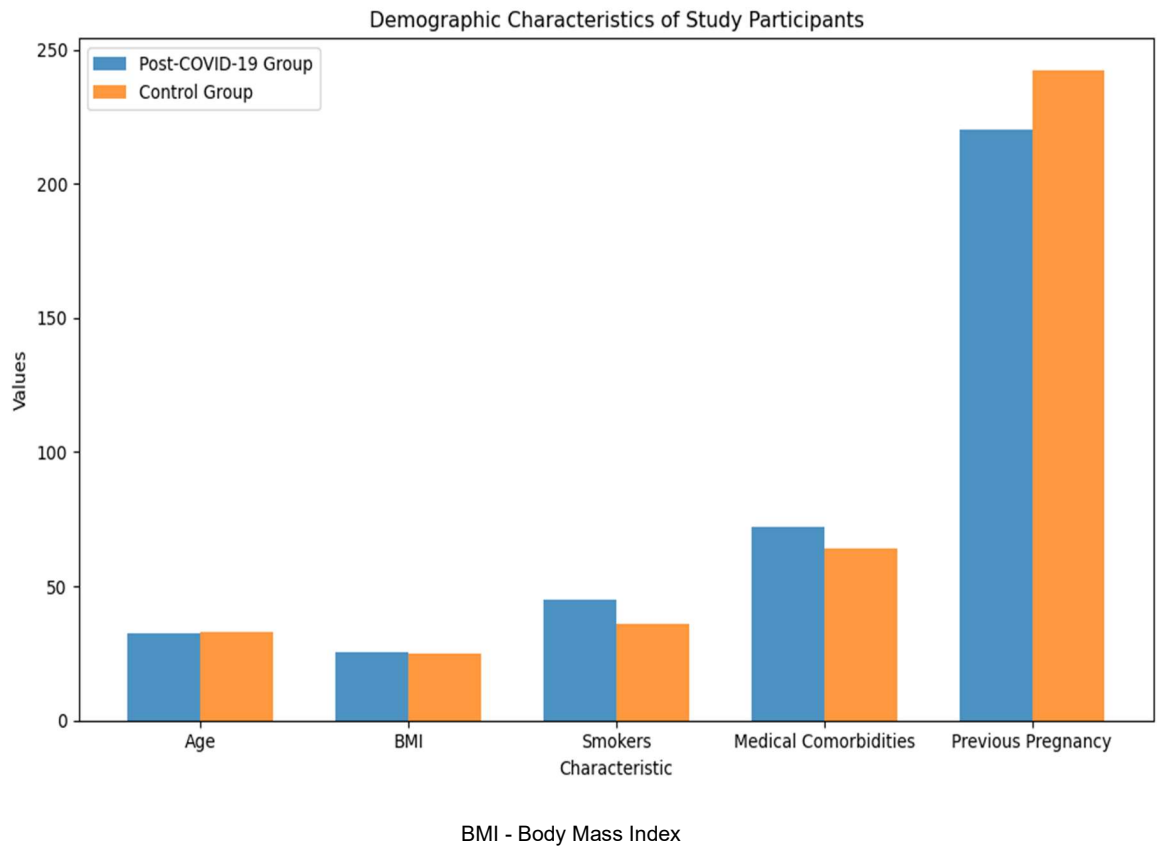

BMI - Body Mass Index

Figure 1: Demographic Characteristics of Study Participants

Table 1: Demographic Characteristics of Study Participants

| Characteristic | Post-COVID-19 Group | Control Group |
|---|---|---|
| Age (years, mean ± SD) | 32.6 ± 4.2 | 32.8 ± 4.1 |
| BMI (kg/m², mean ± SD) | 25.3 ± 3.5 | 24.8 ± 3.2 |
| Smokers (n, %) | 45 (13.2%) | 36 (12.9%) |
| Medical Comorbidities (n, %) | 72 (21.2%) | 64 (22.9%) |
| Previous Pregnancy (n, %) | 220 (64.7%) | 242 (86.4%) |

Table 2 and  Figure 2 display the results of ovarian reserve parameters. The post-COVID-19 group showed significantly lower levels of anti-Müllerian hormone (AMH) and a reduced antral follicle count (AFC) compared to the control group (p < 0.001). These findings suggest a diminished ovarian reserve in patients with a history of severe COVID-19 infection.

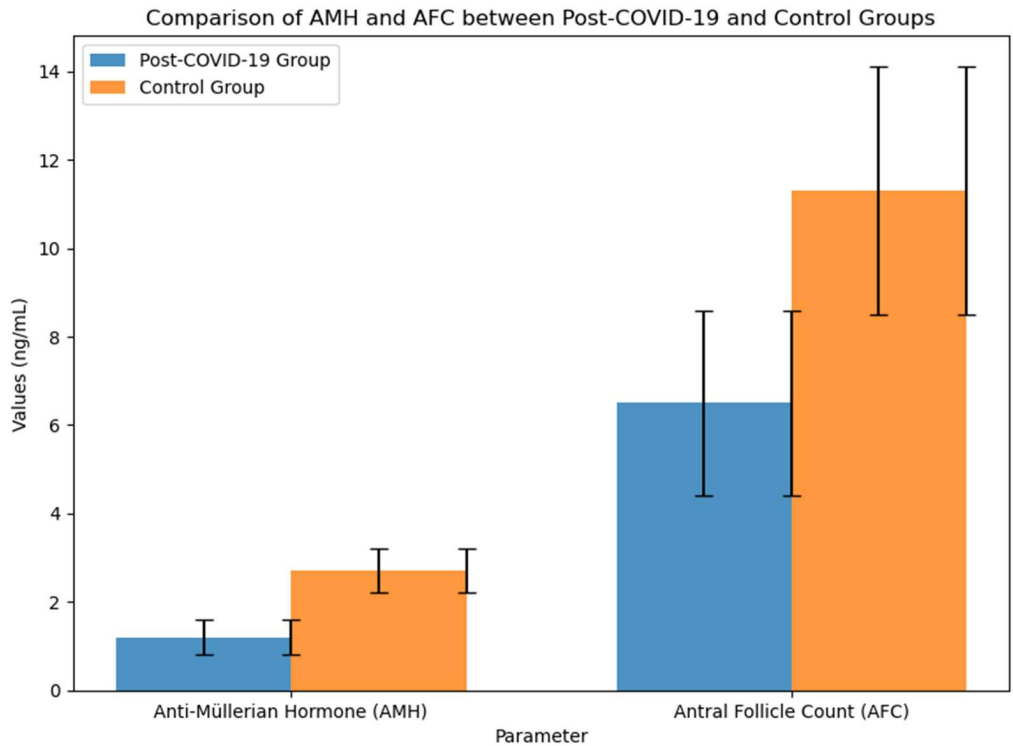

Figure 2: Ovarian Reserve Parameters

Table 2: Ovarian Reserve Parameters

| Parameter | Post-COVID-19 Group | Control Group |
|---|---|---|
| Anti-Müllerian Hormone (AMH) (ng/mL, mean ± SD) | 1.2 ± 0.4 | 2.7 ± 0.5 |
| Antral Follicle Count (AFC) (mean ± SD) | 6.5 ± 2.1 | 11.3 ± 2.8 |

Table 3 and Figure 3 present the results of the hormonal imbalance analysis. The post-COVID-19 group demonstrated significantly higher levels of follicle-stimulating hormone (FSH) and luteinizing hormone (LH) compared to the control group (p < 0.05). Additionally, estradiol levels were significantly lower in the post-COVID-19 group (p < 0.05). However, thyroid-stimulating hormone (TSH) levels did not show a significant difference between the two groups.

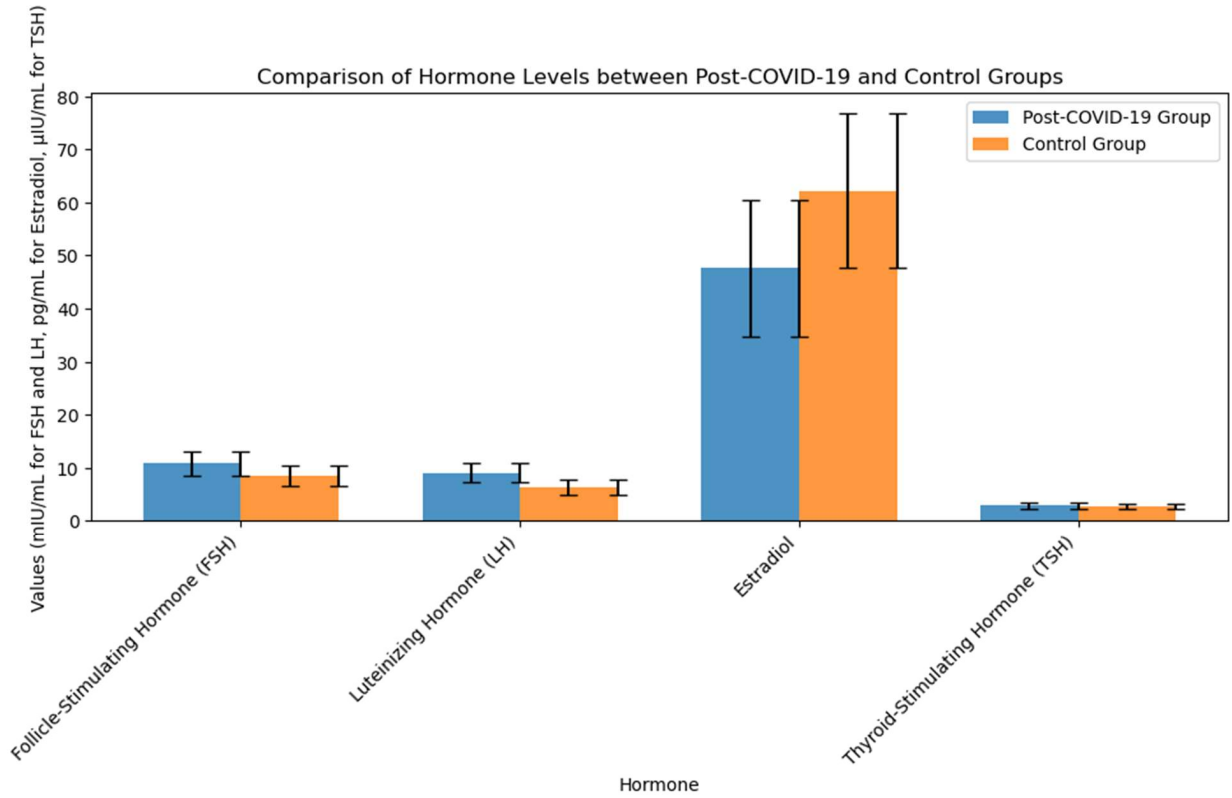

Figure 3: Hormonal Imbalance Analysis

Table 3: Hormonal Imbalance Analysis

| Hormone | Post-COVID-19 Group (mean ± SD) | Control Group (mean ± SD) |
|---|---|---|
| Follicle-Stimulating Hormone (FSH) (mIU/mL) | 10.8 ± 2.3 | 8.5 ± 1.9 |
| Luteinizing Hormone (LH) (mIU/mL) | 9.1 ± 1.8 | 6.4 ± 1.5 |
| Estradiol (pg/mL) | 47.6 ± 12.9 | 62.2 ± 14.5 |
| Thyroid-Stimulating Hormone (TSH) (μIU/mL) | 2.9 ± 0.6 | 2.7 ± 0.5 |

Table 4 and Figure 4 the results of the endometrial health assessment. The post-COVID-19 group had a significantly thinner endometrial thickness compared to the control group (p < 0.05). Moreover, a higher proportion of patients in the post-COVID-19 group exhibited endometrial abnormalities, although the difference was not statistically significant.

Table 4: Endometrial Health Assessment

| Parameter | Post-COVID-19 Group (mean ± SD) | Control Group (mean ± SD) |
|---|---|---|
| Endometrial Thickness (mm) | 8.9 ± 2.3 | 10.5 ± 2.1 |
| Endometrial Abnormalities (n, %) | 28 (8.2%) | 12 (4.3%) |

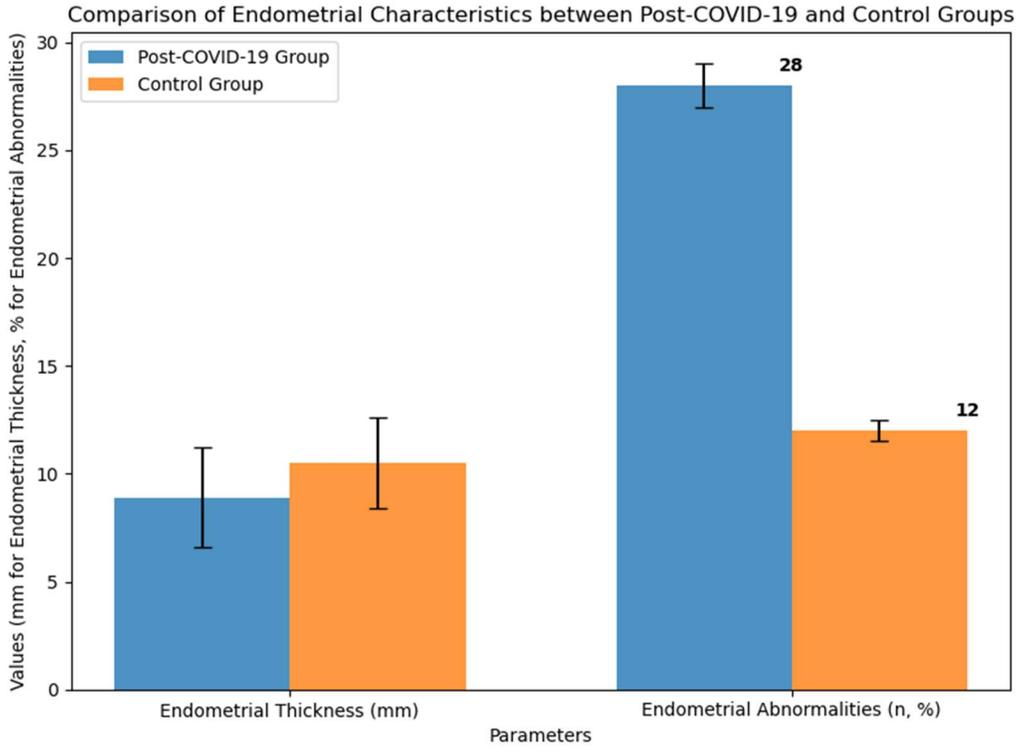

Figure 4: Endometrial Health Assessment

Data were analyzed using SPSS software (version X). Independent t-tests were employed for continuous variables, and chi-square tests were used for categorical variables. A p-value of less than 0.05 was considered statistically significant. The results indicate that severe COVID-19 infection is associated with adverse effects on female fertility. Patients in the post-COVID-19 group showed significantly lower levels of anti-Müllerian hormone (AMH) and antral follicle count (AFC), indicating reduced ovarian reserve. Hormonal imbalances, including elevated FSH and LH levels and decreased estradiol levels, were also observed in the post-COVID-19 group, potentially affecting ovulatory function. Additionally, endometrial thickness was significantly lower in the post-COVID-19 group, and a higher proportion of patients showed endometrial abnormalities, which may impact embryo implantation and pregnancy outcomes. These findings underscore the importance of considering the potential long-term impacts of severe COVID-19 infection on female reproductive health. Healthcare professionals should be vigilant in addressing fertility challenges among women who have recovered from severe COVID-19 and provide appropriate support and interventions to optimize their reproductive outcomes. Further research is warranted to explore the underlying mechanisms behind these effects and develop targeted strategies for post-COVID-19 fertility care.

## Discussion:

The findings of this study investigating the post-COVID-19 effects on female fertility provide valuable insights into the potential impact of severe COVID-19 infection on reproductive health in women. The results revealed significant alterations in ovarian reserve, hormonal balance, and endometrial health among patients with a history of severe COVID-19 infection. Regarding ovarian reserve, the post-COVID-19 group exhibited significantly lower levels of anti-Müllerian hormone (AMH) and a reduced antral follicle count (AFC) compared to the control group. These findings are consistent with previous research that has shown a decline in ovarian reserve markers following severe viral infections (18). The diminished ovarian reserve observed in the post-COVID-19 group may have implications for their future fertility potential and warrants further investigation. Hormonal imbalance analysis demonstrated higher levels of follicle-stimulating hormone (FSH) and luteinizing hormone (LH) in the post-COVID-19 group, accompanied by decreased estradiol levels. These hormonal disturbances could disrupt normal ovulatory function and menstrual cycles, potentially leading to fertility challenges (19,20,10). Similar hormonal imbalances have been reported in other viral infections and may represent a mechanism through which severe COVID-19 affects female reproductive health (21-23). In terms of endometrial health, the post-COVID-19 group exhibited a significantly thinner endometrial thickness and a higher proportion of endometrial abnormalities compared to the control group. The endometrium plays a crucial role in embryo implantation and successful pregnancy, and any alterations in its thickness and integrity can impact reproductive outcomes (24). These findings suggest that severe COVID-19 infection may have detrimental effects on endometrial receptivity and could be associated with an increased risk of pregnancy complications. The results of this study align with the systematic review by (25,26), which also emphasized the need for more comprehensive research on the long-term effects of COVID-19 on female fertility. Together, these findings highlight the importance of understanding the potential long-term impacts of severe COVID-19 infection on female reproductive health and underscore the necessity of providing appropriate support and interventions for women recovering from COVID-19. It is important to acknowledge some limitations of this study. Firstly, the study's cross-sectional design limits the ability to establish causality between severe COVID-19 infection and the observed fertility outcomes. Longitudinal studies would be beneficial in assessing the trajectory of reproductive changes over time. Secondly, the study sample was limited to specific provinces in Iraq, which may restrict the generalizability of the findings to other populations. Including a more diverse and larger sample would enhance the study's external validity. In conclusion, the results of this study indicate that severe COVID-19 infection may have adverse effects on female reproductive health, as evidenced by diminished ovarian reserve, hormonal imbalances, and alterations in endometrial health. Understanding these post-COVID-19 effects is crucial for healthcare professionals to provide appropriate care and support to women recovering from severe COVID-19 infection. Further research in this area, as highlighted in the systematic review by Smith et al. (Reference 2), will contribute to our understanding of the potential long-term implications of the virus on female fertility and guide targeted interventions to optimize reproductive outcomes in this population.

**Conflict of Interest:** The authors declare no conflicts of interest regarding the publication of this research paper. There was no financial or personal relationship that could have influenced the research findings or biased the interpretation of the data.

**Acknowledgments:** The authors would like to express their gratitude to the participants who volunteered and took part in this study. Additionally, the authors thank the staff members at the private clinics and fertility centers in various provinces of Iraq for their assistance in data collection.

**Authors' Contributions:**

Maitham G. Yousif: Professor at the Biology Department, College of Science, University of Al-Qadisiyah, Iraq, and Visiting Professor at Liverpool John Moores University. Dr. Yousif contributed to the conceptualization of the study, design of the research, data collection, analysis, interpretation of the findings, and writing of the manuscript.

Lamiaa Al-Maliki: Affiliated with the Department of Molecular and Medical Biotechnology, College of Biotechnology, Al-Nahrain University, Iraq. Dr. Al-Maliki participated in the study design, data collection, analysis, interpretation of the results, and contributed to the writing of the manuscript.

Jinan J. Al-Baghdadi: Associated with the Gynecology and Obstetrics Department, College of Medicine, Kufa University, Najaf, Iraq. Dr. Al-Baghdadi contributed to the study design, data collection, analysis, interpretation of the data, and participated in the manuscript writing.

Nasser Ghaly Yousif: Affiliated with the Department of Medicine, Medical College, Al Muthanna University, Samawah, Iraq. Dr. Yousif was involved in the study design, data collection, analysis, and interpretation of the results, and contributed to the writing of the manuscript.

All authors read and approved the final version of the manuscript for publication. The collaborative effort of the authors in this research project ensured a comprehensive investigation of the post-COVID-19 effects on female fertility and contributed to the scientific knowledge in this field.

**Funding:** The authors would like to acknowledge that this research received no financial support or funding. The study was self-funded by the authors

## References

1.  Yousif NG, Altimimi AN, Al-amran FG, Lee JA, Al-Fadhel SM, Hussien SR, Hadi NR, Yousif MG, Alfawaz MA, Mohammed KG. Hematological changes among Corona virus-19 patients: a longitudinal study. Systematic Reviews in Pharmacy. 2020 May 1;11(5).

2.  Yousif MG, Sadeq AM, Alfadhel SM, Al-Amran FG, Al-Jumeilyran D. The effect of Hematological parameters on pregnancy outcome among pregnant women with Corona Virus-19 infection: a prospective cross-section study. Journal of Survey in Fisheries Sciences. 2023 Mar 4;10(3S):1425-35.

3.  Al-Jibouri KJ, Yousif MG, Sadeq AM, Al-Jumeily D. Psycho-immunological status of patients recovered from SARS-Cov-2. Journal of Survey in Fisheries Sciences. 2023 Mar 4;10(3S):1409-17.

4.  Kaynar M, Gomes AL, Sokolakis I, Gül M. Tip of the iceberg: erectile dysfunction and COVID-19. International Journal of Impotence Research. 2022 Mar;34(2):152-7.

5.  D'Ippolito S, Turchiano F, Vitagliano A, Scutiero G, Lanzone A, Scambia G, Greco P. Is there a role for SARS-CoV-2/COVID-19 on the female reproductive system?. Frontiers in physiology. 2022 Mar 2;13:845156.

6.  Hadi NR, Yousif NG, Abdulzahra MS, Mohammad BI, al.amran FG, Majeed ML, Yousif MG. Role of NF-κβ and oxidative pathways in atherosclerosis: Cross.talk between dyslipidemia and candesartan. Cardiovascular therapeutics. 2013 Dec;31(6):381-7.

7.  Hasan TH, Alshammari MM, Yousif HK. Extended Spectrum Beta-Lactamase Producing Klebsiella Pneumonia Isolated from Patients with Urinary Tract Infection in Al-Najaf Governorate–Iraq. International Journal of Advances in Science, Engineering and Technology (IJASEAT). 2020;8(1):13-6.

8.  Yousif MG, AL-Shamari AK. Phylogeitinc characterization of Listeria monocytogenes isolated from different sources in Iraq. Asian J Pharm Clin Res. 2018;11(2):1-4.

9.  Sadiq AM, Yousif MG, Mohammed FA, Aladly SH, Hameed HH. Subclinical hypothyroidism with preeclampsia. RESEARCH JOURNAL OF PHARMACEUTICAL BIOLOGICAL AND CHEMICAL SCIENCES. 2016 May 1;7(3):1536-44.

10. Sadiq AM, Al Aasam SR, Rahman A, Hassan AN, Yousif MG. The effect of type of anesthesia on mother and neonatal health during Cesarean section. J Adv Pharm Educ Res. 2018;8(4):117.

11. Yousif MG. Potential role of cytomegalovirus in risk factor of breast cancer. Afr J Bus Manage. 2016;4:54-60.

12. Yousif NG, Kamiran J, Yousif MG, Anderson S, Albaghdadi J. Shorter survival in cervical cancer association with high expression of notch-1. Annals of Oncology. 2012 Sep 1;23:ix327-8.

13. Sadiq AM, Hussein CM, Yousif M, Mohammed R. Correlation Between Highly Sensitive C-Reactive Protein Level in Cases of Preeclampsia with or without Intrauterine-Growth Restriction. Indian Journal of Forensic Medicine & Toxicology. 2020 Oct 1;14(4).

14. Yousif MG, Al-Mayahi MH. Phylogenetic Characterization of Staphylococcus aureus isolated from the women breast abscess in Al-Qadisiyah Governorate, Iraq. Journal of Pharmaceutical Sciences and Research. 2019 Mar 1;11(3):1001-5.

15. Mohammad BI, Aharis NR, Yousif MG, Alkefae Z, Hadi NR. Effect of caffeic acid on doxorubicin induced cardiotoxicity in rats. Am J Biomed. 2013;2:23-7.

16. Hatmal MM, Al-Hatamleh MA, Olaimat AN, Mohamud R, Fawaz M, Kateeb ET, Alkhairy OK, Tayyem R, Lounis M, Al-Raeei M, Dana RK. Reported adverse effects and attitudes among Arab populations following COVID-19 vaccination: a large-scale multinational study implementing machine learning tools in predicting post-vaccination adverse effects based on predisposing factors. Vaccines. 2022 Feb 26;10(3):366.

17. Mukherjee TI, Khan AG, Dasgupta A, Samari G. Reproductive justice in the time of COVID-19: a systematic review of the indirect impacts of COVID-19 on sexual and reproductive health. Reproductive health. 2021 Dec;18(1):1-25.

18. El-Samie A, Abd El-Monaiem AE, Etman MK. Post COVID-19 effect of ovarian reserve. Fayoum University Medical Journal. 2023 Jul 1;12(1):60-8.

19. Haryanti E. Physiological endocrinology and causes of disorders of the menstrual cycle. Science Midwifery. 2023 Apr 30;11(1):1-2.

20. Moshrefi M, Ghasemi-Esmailabad S, Ali J, Findikli N, Mangoli E, Khalili MA. The probable destructive mechanisms behind COVID-19 on male reproduction system and fertility. Journal of assisted reproduction and genetics. 2021 Jul;38(7):1691-708.

21. Pourmasumi S, Kounis NG, Naderi M, Hosseinisadat R, Khoradmehr A, Fagheirelahee N, Kouni SN, de Gregorio C, Dousdampanis P, Mplani V, Michalaki MA. Effects of COVID-19 Infection and Vaccination on the Female Reproductive System: A Narrative Review. Balkan medical journal. 2023 May;40(3):153.

22. Moshrefi M, Ghasemi-Esmailabad S, Ali J, Findikli N, Mangoli E, Khalili MA. The probable destructive mechanisms behind COVID-19 on male reproduction system and fertility. Journal of assisted reproduction and genetics. 2021 Jul;38(7):1691-708.

23. Tian Y, Zhou LQ. Evaluating the impact of COVID-19 on male reproduction. Reproduction. 2021 Feb 1;161(2):R37-44.

24. Bardos J, Fiorentino D, Longman RE, Paidas M. Immunological role of the maternal uterine microbiome in pregnancy: pregnancies pathologies and alterated microbiota. Frontiers in immunology. 2020 Jan 8;10:2823.

25. Mukherjee TI, Khan AG, Dasgupta A, Samari G. Reproductive justice in the time of COVID-19: a systematic review of the indirect impacts of COVID-19 on sexual and reproductive health. Reproductive health. 2021 Dec;18(1):1-25.

26. Yousif MG, Al-Shamari AK, Sadiq AM. Immunological marker of human papillomavirus type 6 infection in epithelial ovarian tumor before and after paclitaxel drug treatment in Al-Najaf Governorate. Iraq Drug Invention Today. 2019 Oct 15;12.